\documentclass[letterpaper,twocolumn,10pt]{article}
\usepackage{usenix2019_v3}
\usepackage{comment}

\usepackage{tikz}
\usepackage{placeins}
\usepackage{algorithm}
\usepackage{listings}
\usepackage{xcolor}
\usepackage{yquant/yquant} 
\usepackage[most]{tcolorbox}
\usepackage[noend]{algpseudocode} 

\newcommand{\tool}{{\small{\textsc{HALO}}}\xspace}

\usepackage{xspace}
\usepackage{braket} 

\usepackage{amsmath}
\usepackage{multirow}
\usepackage{makecell}
\usepackage{filecontents}

\begin{document}

\date{}

\title{\Large \bf HALO: A Fine-Grained Resource Sharing Quantum Operating System}

\author{
{\rm John Zhuoyang\ Ye}\\
University of California, Los Angeles
\and
{\rm Jiyuan Wang}\\
Tulane University
\and
{\rm Yifan Qiao}\\
University of California, Berkeley
\and
{\rm Jens Palsberg}\\
University of California, Los Angeles
} 

\maketitle

\begin{abstract}
As quantum computing enters the cloud era, thousands of users must share access to a small number of quantum processors. Users need to wait minutes to days to start their jobs, which only takes a few seconds for execution. Current quantum cloud platforms employ a \emph{fair-share} scheduler, as there is no way to multiplex a quantum computer among multiple programs at the same time, leaving many qubits idle and significantly under-utilizing the hardware. This imbalance between high user demand and scarce quantum resources has become a key barrier to scalable and cost-effective quantum computing.

We present \tool{}, the first quantum operating system design that supports fine-grained resource-sharing.  \tool introduces two complementary mechanisms.
First, a hardware-aware qubit-sharing algorithm that places shared helper qubits on regions of the quantum computer that minimize routing overhead and avoid cross-talk noise between different users' processes. Second, a shot-adaptive scheduler that allocates execution windows according to each job’s sampling requirements, improving throughput and reducing latency. Together, these mechanisms transform the way quantum hardware is scheduled and achieve more fine-grained parallelism.

We evaluate \tool on the IBM Torino quantum computer on helper qubit intense benchmarks. Compared to state-of-the-art systems such as HyperQ, \tool improves overall hardware utilization by up to 2.44$\times$, increasing throughput by 4.44$\times$, and maintains fidelity loss within 33\%, demonstrating the practicality of resource-sharing in quantum computing.
\end{abstract}


\section{Introduction}
\label{sec:Intro}

Quantum computing has emerged as one of the most promising paradigms for solving problems that are intractable on classical machines, offering exponential speed-ups in applications such as cryptography, optimization, and materials simulation\cite{Shor,grover1996fastquantummechanicalalgorithm,arute2019quantum,Hao_2025}. Over the past decade, rapid advances in superconducting and trapped-ion hardwares have led to the deployment of cloud-accessible quantum computers by major vendors such as IBM~\cite{steffen2011quantum,abughanem2024ibm}, Google~\cite{google2023suppressing,arute2019quantum}, and Rigetti~\cite{motta2020determining,zeng2017first}. 

Despite this progress, the quantum computing ecosystem faces a critical hardware bottleneck. The number of available quantum devices is still extremely small, while the global user base continues to grow rapidly. On public platforms such as IBM Quantum, thousands of users share access to a handful of machines, often leading to queues with hundreds or thousands of pending jobs~\cite{ibm-busy}. This mismatch between high user demand and limited hardware capacity poses a serious barrier to scalable and cost-effective quantum computing.

Existing cloud platforms primarily adopt an exclusive-use model, where each submitted job monopolizes the entire quantum device—even if it requires only a small fraction of the available qubits or execution time~\cite{ibm-fairshare,braket-priority,braket-reservations}. As a result, qubit utilization remains low, queue times are long, and the overall throughput of the system is severely constrained. A typical IBMQ cloud service has a queue with up to thousands of pending jobs, and users have to wait for hours to get their results back~\cite{tao2025quantum,ibm-busy}. 

To alleviate this bottleneck, we argue that the quantum computing ecosystem urgently needs a \emph{system-level framework} for efficient and flexible resource utilization. Recent efforts toward this goal, such as quantum operating systems (QOS)~\cite{giortamis2024qos} and quantum virtual machine (QVM) frameworks~\cite{tao2025quantum}, represent promising first steps. QOS focuses on providing abstractions for program lifecycle management, job submission, and hardware access control, while QVM systems aim to virtualize quantum devices to support concurrent program execution and improve user-level responsiveness. However, despite these advances, existing solutions remain limited in the following two ways.  

\begin{figure}[h!]
    \centering
    \includegraphics[width=0.35\textwidth]{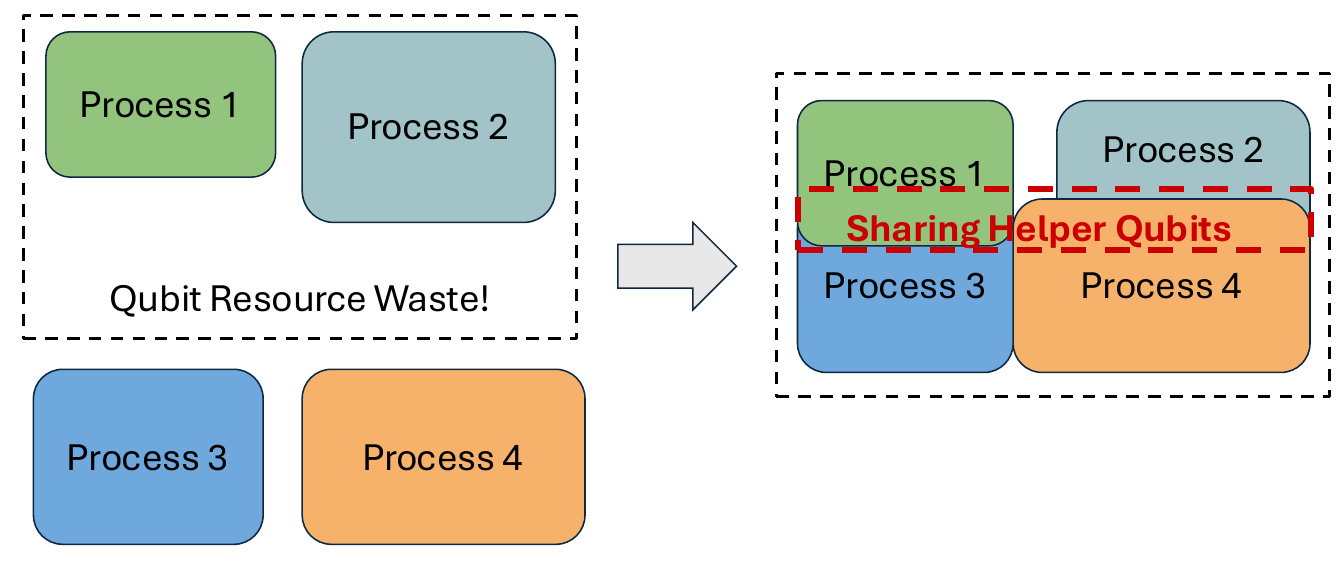}
    \caption{Sharing helper qubits can in principle increase system throughput for applications need a lot of helper qubits.}
    \label{fig:share}
\end{figure}

\noindent\textbf{Limitation 1: Unaware of helper qubit sharing opportunities.} Many quantum algorithms rely on \emph{helper qubits} (also known as \emph{ancilla} qubits) that are borrowed temporarily during computation and released afterward\cite{brown2011ancilla,anders2010ancilla}. However, existing systems either time-share the entire quantum computer and dedicate all qubits to a single program at a time (e.g., QOS), or statically partition qubits across concurrent programs. They do not exploit the opportunity to reuse these qubits across concurrent processes, as illustrated in Figure \ref{fig:share}.

    
\noindent\textbf{Limitation 2: Shot-unaware scheduling.} Current schedulers in previous work such as HyperQ\cite{tao2025quantum} and QOS\cite{giortamis2024qos} assume that all processes need same amount of \emph{shots} (repeated circuit executions for statistical accuracy). However, real quantum applications vary in shot demand. Some fault-tolerant algorithm, such as Shor and Grover's algorithm, needs only a few shots~\cite{Shor,grover1996fastquantummechanicalalgorithm}. Some other recent NISQ algorithms, such as QAOA and random circuit sampling algorithm, need thousands or even millions of shots\cite{arute2019quantum,Hao_2025}. As a result, a scheduler that ignores differences in shot requirements cannot properly handle users' applications in general and fully optimize space-time utilization.  


While space sharing and time sharing offer substantial opportunities for improving device utilization, enabling these capabilities in practice introduces several technical challenges.

\noindent\textbf{Challenge 1: Qubit sharing amplifies noise.}
Sharing helper qubits across concurrently executing processes imposes strict requirements on how each program’s data qubits are mapped onto the hardware\cite{ShareAncilla}. To safely co-locate multiple circuits, the system must place their data qubits sufficiently close to shared helper qubits to avoid excessive SWAP insertion while also ensuring they remain adequately separated to limit crosstalk and unintended entanglement\cite{zhao2022quantumCrosstalk}. These constraints couple the mapping decisions of all participating processes, turning what is typically a single-circuit optimization into a joint placement problem with larger search space. 

\noindent\textbf{Challenge 2: Shot-aware scheduling requires coordinated temporal multiplexing.} The system must evaluate combinations of jobs whose execution windows overlap in time, respect hardware-level concurrency constraints, and avoid unfair allocation patterns. Moreover, scheduling decisions must account for different shot demands so that long-shot jobs do not monopolize the hardware and short-shot jobs do not incur unnecessary waiting.


\begin{table}[h!]
\centering
\small
\renewcommand{\arraystretch}{1}
\begin{tabularx}{\columnwidth}{l *{4}{C}}
\hline
\textbf{Feature} 
& \textbf{IBM\cite{ibm-fairshare}} 
& \textbf{QVM\cite{tao2025quantum}} 
& \textbf{QOS\cite{giortamis2024qos}} 
& \textbf{HALO} \\
\hline
Shots aware             & \gcmark & \rxmark & \rxmark & \gcmark(\S \ref{sec:shot-aware-schedul}) \\
Space multiplexing      & \rxmark & \gcmark & \gcmark & \gcmark(\S \ref{subsec:SpaceManager})  \\
Helper qubit sharing    & \rxmark & \rxmark & \rxmark & \gcmark(\S \ref{subsec:HelperScheduler}) \\
\hline
\end{tabularx}

\caption{Feature comparison among current quantum systems with schedulers. \tool{} support all features by methods in this paper. }
\label{tab:Features}
\end{table}

We develop \tool{} to address these two challenges and to advance the state of quantum resource scheduling beyond existing approaches summarized in Table \ref{tab:Features}.IBM Quantum can schedule user processes with different shots but is unaware of helper qubits and doesn't support space multiplexing. Both QOS and HyperQ support space multiplexing but cannot schedule processes with different shots and is unaware of helper qubit sharing opportunities.

\tool{} is the first quantum OS design which combine all three scheduling features.  More specifically, \tool{} solves the challenge by using a fine-grained resource sharing mechanism which enables both dynamic space sharing (qubit-level concurrency) and time sharing (shot-aware execution) on real quantum hardware. On the spatial side, \tool supports users to dynamically share the helper qubits between different processes, and minimize the noise with a cost function to jointly optimize qubit placement, routing cost, and noise isolation when multiple programs co-execute on the same device. On the temporal side, \tool incorporates a shot-adaptive scheduler that allocates execution windows according to each job’s shots count and circuit depth. Together, these mechanisms allow \tool to substantially increase quantum hardware utilization while maintaining correctness. We evaluate \tool on three helper qubit intensive benchmarks: the stabilizer measurement circuits, classical arithmetic circuits, and the multi-qubit controlled gates. The main contributions of our paper:

\begin{itemize}
    \item 
    We design \tool{}, the first fine-grained resource sharing design that enables multiple quantum processes to safely co-execute on the same quantum device through coordinated space sharing (sharing helper qubits) and time sharing (shot-aware scheduling).
    \item We develop a hardware-aware multi-process mapping model, including a cost function that accounts for the placement of data-qubit and helper-qubit zone, as well as cross-talk isolation. Our cost functions can explain the fidelity degradation when throughput increase from experiments(\S \ref{subsec:RQ4}). In addition, we design a shot-aware scheduler that balances processes with different shot requirements to reduce the idle time and improve the batch utilization(\S \ref{sec:shot-aware-schedul}).
    \item \tool{} outperforms the state-of-art work by $2.44\times $ on utilization, $4.44 \times$ on throughput, while only sacrifice a moderate fidelity loss within $33\%$(\S \ref{subsec:RQ1}).
\end{itemize}

\section{Background}
\label{sec:Background}

\paragraph{Quantum Computing.}
Quantum computing has emerged as a promising computational paradigm capable of solving certain problems exponentially faster than classical computers\cite{Shor,daley2022practical}. These capabilities stem from inherently quantum phenomena such as entanglement, which allow quantum programs to explore large computational spaces more efficiently than classical algorithms\cite{entanglement}. As quantum hardware continues to improve, cloud-accessible quantum processors have become increasingly available, allowing researchers across academia and industry to execute real quantum programs\cite{abughanem2024ibm}. However, the limited number of devices and their constrained qubit resources create significant challenges for scaling quantum workloads, motivating new system-level approaches for efficient resource management\cite{ibm-busy}.

\paragraph{Data Qubits and Helper Qubits.} In quantum computation, there are two types of qubits: data qubits, and helper qubits (ancilla qubits)\cite{conditional,brown2011ancilla,anders2010ancilla,nielsen2010quantum}. Data qubits are qubits that store and manipulate key quantum information. Helper qubits, on the other hand, serve an auxiliary role and only have a rather short lifetime. Helper qubits are essential for almost all quantum algorithms at scale.
For example, they are used to help make the compilation reversible or more efficient\cite{nielsen2010quantum,brown2011ancilla,anders2010ancilla,jayashree2016ancilla}. In fault-tolerant quantum computation, fresh helper qubits are prepared and used to detect error syndromes\cite{shor1996fault}. Also, lots of helper qubits are necessary in fault-tolerant protocols such as magic state distillation and lattice surgery\cite{bravyi2012magic,Litinski_2019}. One thing is common for helper qubits in all quantum computation: after their jobs complete, they are unentangled with data qubits and can be seen as garbage. Thus, we are able to reuse helper qubits in future computation or even in other processes\cite{jayashree2016ancilla,ShareAncilla}. Given the common reusability nature of helper qubits, we argue in this research that helper qubits should be seen as a unique type of resources and be managed by the operating system kernel.





\paragraph{Entanglement.}
Entanglement is a uniquely quantum phenomenon in which the joint state of multiple qubits cannot be expressed as a product of individual states\cite{entanglement,gisin1998bell}. While entanglement is indispensable for many quantum algorithms, it imposes strict isolation requirements in a multi-tenant quantum system like \tool{}. Different processes must be unentangled throughout computation. For example, consider two quantum programs $P_A$ and $P_B$ that reuse
the same helper qubit $q$ at different points in time. If $q$ is not properly reset after $P_A$ finishes its use, quantum entanglement may cause $P_B$'s operations to alter the quantum state of $P_A$, leading to incorrect outputs.Thus, before reassigning a helper qubit, \tool{} must ensure that it is disentangled and reset to initial state $\ket{0}$. In other words, \tool{} must carefully control qubit sharing to \emph{prevent any cross-process entanglement}, ensuring that each job’s quantum state evolves independently even while sharing hardware resources.

\paragraph{Quantum process and crosstalk.}
A quantum program consists of a quantum circuit—a sequence of unitary operations, measurements, and reset instructions—along with a specified number of \emph{shots}, which determines how many times the circuit is repeatedly executed and measured to obtain a classical result\cite{sinanan2024single,nielsen2010quantum}. In a multi-tenant quantum system, multiple processes may be mapped onto different regions of the hardware simultaneously\cite{giortamis2024qos,tao2025quantum,ShareAncilla}. However, physical qubits are not perfectly isolated: superconducting devices exhibit \emph{crosstalk}, where control pulses or two-qubit operations applied to one qubit unintentionally perturb the state or error rates of nearby qubits\cite{zhao2022quantumCrosstalk}. Crosstalk can degrade fidelity even when processes do not directly interact, and its effects become more severe when processes are placed too closely or share routing paths. Therefore, \tool{} must assign qubits and schedule operations in a manner that minimizes interference between processes. 

\paragraph{Quantum hardware and layout mapping.}
There are already many different physical platforms that support quantum computation, such as superconducting\cite{clarke2008superconducting}, trapped ions\cite{bruzewicz2019trapped}, or neutral atoms\cite{evered2023high}. To compile any abstract quantum circuit to all these hardware, finding a good layout mapping from abstract qubit to real qubits is a common problem, due to hardware layout constraints\cite{tan2020optimal}, because to connect two remote qubits, we have to add extra noisy routing CNOT gates. In this paper, we consider a superconducting quantum computer with fixed hardware layout graph. 




\section{Design}
\label{sec:Des}
\begin{figure*}[htbp]
    \centering
    \includegraphics[width=0.7\textwidth]{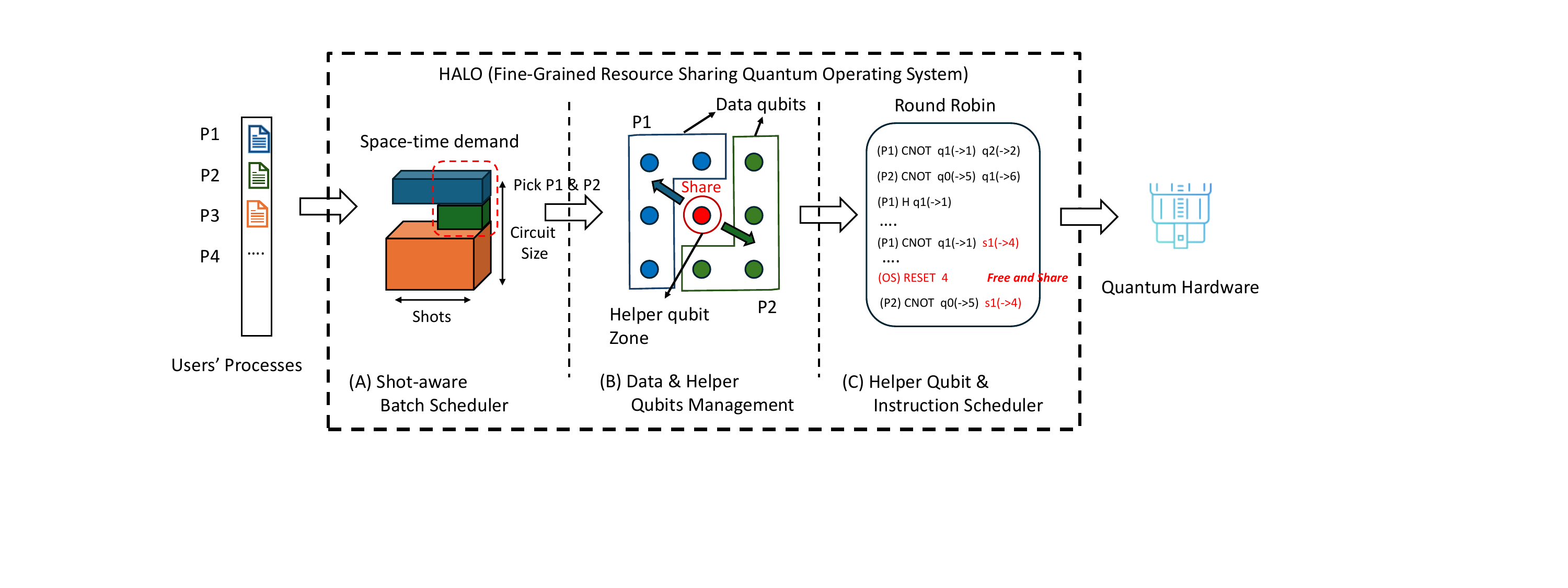}
    \caption{The work flow diagram of HALO scheduler in our resource sharing quantum operating system. It takes user's processes as input and send scheduled quantum gate instructions to the quantum hardwares.}
    \label{fig:HALO}
\end{figure*}

\subsection{Design Priciples}
\label{sec:DesPrin}
\paragraph{Virtualization of layout, shots and helper qubits.}
Inspired from the recent seminal work of QVM \cite{tao2025quantum} and QOS\cite{giortamis2024qos}, we virtualize the space-time resource to facilitate the management and scheduling process\cite{denning1970virtual}. Different from HyperQ, we separate the virtualization of helper qubits and data qubits. In our design, layout of all qubits in the processes are virtualized, and the layout mapping are completely mananaged by scheduler in QS. Also, we virtualize the shot of each process as an essential time resource.  

\paragraph{Fine-grained resource sharing.}
A central objective of scheduling in \tool{} is to maximize overall system throughput by increasing the number of programs executed concurrently. To achieve this, the system must maximize the QPU usage qubit-sharing opportunities when many processes want to access the hardware simultaneously. During the space-time scheduling, two quantum processes are expected to share some of the hardware qubits as their helper qubits at different times. Moreover, such opportunities such be exploited as much as possible to maximize the utility of qubit space in hardware. Unlike prior systems such as HyperQ~\cite{tao2025quantum}, which rely on manually specified virtual layouts, \tool{} performs \emph{automated fine-grained layout synthesis} for concurrently executing circuits. The mapping of all virtual data qubits is a function that accounts for real-device connectivity. At the core of the mapping function is a new cost function that considers multiple constraints unique to multi-tenant quantum execution—routing distance, inter-process crosstalk, and helper-qubit reuse opportunities. This unified objective enables \tool{} to quantify the safety and benefit of sharing decisions, reducing the extra error caused by noisy routing circuits.

\paragraph{Flexible tradeoff between utilization and fidelity.}
We provide users with a flexible tradeoff between qubit utilization and fidelity. Through a global system configuration, users can choose to sacrifice fidelity for higher throughput or, conversely, reduce throughput to gain higher fidelity. We believe such flexibility is essential for near-term quantum operating systems, where hardware resources are scarce, and user requirements are highly diverse\cite{ibm-busy,giortamis2024qos}.


\paragraph{Isolation across processes.}
In the qubit layout mapping across many processes, the scheduler has to make sure that each process is isolated to prevent unintended interference\cite{tao2025quantum,giortamis2024qos}. First, each process’s data qubits are mapped to \emph{disjoint physical regions}, preventing accidental entanglement or crosstalk with other processes. Our placement algorithm explicitly avoids qubit assignments that violate spatial separation constraints by \textit{inter-process crosstalk cost} discussed in Section 3.2. Second, for helper qubits that are shared across processes, \tool{} enforces a mandatory \texttt{reset} operation before reuse, ensuring the qubits are fully disentangled from the previous workload. Together, these mechanisms ensure that each process evolves independently, even while the hardware resources are being multiplexed for improved utilization.

\subsection{Kernel System Call Interfaces}
\label{subsec:KernelCall}

\begin{table}[t!]
\centering
\small
\caption{System calls for the resource sharing quantum operating system}
\label{tab:tgate-syscalls}
\begin{tabular}{l l}
\hline
System Call & Description \\
\hline
\texttt{q=alloc\_data(x)} &
\makecell[{{p{4cm}}}]{Allocate $x$ data qubits.}\\

\texttt{s=alloc\_helper(y)} &
\makecell[{{p{4cm}}}]{Allocate $y$ helper qubits.}\\

\texttt{deallocate\_data(q)} &
\makecell[{{p{4cm}}}]{Release the  data-qubit space.}\\

\texttt{deallocate\_helper(si)} &
\makecell[{{p{4cm}}}]{Release the helper-qubit $s_i$.}\\

\texttt{set\_shot(x)} &
\makecell[{{p{4cm}}}]{Set number of shots as $x$.}\\
\hline
\end{tabular}
\label{tab:Interfaces}
\end{table}
\tool{} virtualizes data qubits, helper qubits and shots.  \tool{} has five system interfaces provided to processes by the kernel, as listed in table \ref{tab:Interfaces}. A process should tell the kernel the total number of data qubits and helper qubits, as well as the number of shots it needs. A process asks the kernel to allocate these resources for them by system calls before scheduling starts. A process accesses a virtual data qubit by \texttt{qi}, and a virtual helper qubit by\texttt{si}, where i is the unique index in the one-dimensional virtual space.  All classical measurement results are stored in a one-dimensional virtual classical space labeled as $c_0,c_1,c_2,\cdots$. The scheduler sends the instructions to the hardware after allocation.

Unlike HyperQ \cite{tao2025quantum}, qubit layouts are completely managed by the space manager in \tool{}. As a result, programmers are shielded from hardware layout concerns. Thus, our interface design is adaptable to different quantum hardware and layouts, in principle.

\subsection{\tool{} Overview}
\label{subsec:Tool}


We introduce HALO, a space-time quantum operating system that performs multi-process management and scheduling as shown in Figure~\ref{fig:HALO}. We give an overview of the entire fine-grained scheduling loop. \tool{} coordinates concurrent execution through a three-stage scheduling pipeline:  

1. \textbf{Shot-aware batch scheduling (\S \ref{sec:shot-aware-schedul}).} In the first stage, \tool{} selects a compatible batch of processes from the waiting queue based on space requirements and concurrency potential. After this step, \tool{} decides to run these quantum processes in the selected batch on the hardware for the same number of shots. 
2. \textbf{Static data-qubit placement (\S \ref{subsec:SpaceManager}).} In this stage, \tool{} computes an optimized hardware-aware data qubit layout for all processes in the selected batch of the previous step. \tool{} optimizes a cost function by simulated annealing with some heuristic. All the rest of the unassigned hardware qubits form the helper-qubit zone.
3. \textbf{Dynamic helper-qubit routing and instruction scheduling (\S \ref{subsec:HelperScheduler}).} In the last stage, \tool{} decides on the final instruction order and dynamic mapping of helper qubits for all batch processes. \tool{} uses a round robin algorithm for the instruction order, so every process has a fair chance to compete for the public helper qubit zones. \tool{} greedily assigns the nearest available helper qubits for all instructions.  \tool{} also inserts necessary resets for safe helper qubit reuse, this make sure that all processes are unentangled even if they share helper qubits.









\subsection{Shot-aware process batch scheduler}
\label{sec:shot-aware-schedul}


 In the first stage, \tool{} balance the space-time volume of each process batch to improve the throughput of a single QPU. After the space manager determines a layout
for each process, each process $P$ has $q_p^d$ data qubits and $q_p^h$ helper qubits. And need to execute $s_p$ shots by calling interface \texttt{set\_shot(sp)} listed in Table \ref{tab:Interfaces}. We also denote the circuit depth of the process as $d_p$. 

Let the number of physical qubits on the device as $Q_{\text{tot}}$. \tool{} introduces a tunable system parameter $0 < \lambda < 1$, the \emph{data-qubit occupancy ratio}, which controls the aggressiveness of spatial sharing. The \tool{} forms a batch $B$ of processes $p_{i}$ until the total number of required \emph{data} qubits reaches $\lambda Q_{\text{tot}}$.

Since IBMQ doesn't report the active time for each qubit\cite{tao2025quantum}, we estimate the time of a process in terms of its depth of circuit $\times$ shot count. For a process $P$ with $q_p^d$ data qubits, $q_h^p$ helper qubits, circuit depth $d_p$, and the remaining shot count $s_p$, the \emph{total depth demand} is
$D_p = s_p \cdot d_p$,
and the \emph{space--time demand} (work volume) is
$
W_p = q^d_p \cdot D_p = q^d_p \cdot d_p \cdot s_p
$
, where $Q_p$ is the number of data qubits it needs. When \tool{} schedules a batch $B$ of processes that execute the same number of shots $s_B$, the batch \emph{makespan} is
$
D_B = s_B \cdot \max_{p \in B}(d_p),
$
and the total \emph{space--time capacity} for the batch is
$
\mathrm{Cap}_B = Q_{\text{tot}} \cdot D_B.
$
The batch’s \emph{useful work} is computed as:
$
\mathrm{Work}_B = \sum_{p \in B} q^d_p \cdot d_p \cdot s_B.
$\; \tool{} optimize batch formation using the following heuristic:
\begin{enumerate}
    \item \textbf{Shot and depth aware priority:} Figure~\ref{fig:HALO} shows that resource inefficiency primarily arises from the misalignment between circuit depth and shot count in the process queue. To mitigate this fragmentation, processes are reordered first by jointly considering their shot and depth demands. Specifically, processes with smaller total depth demand $D_p = s_p d_p$ are assigned higher priority. $s_p$ is the remaining shot count and will be updated after batch execution. Similar priority heuristics are used in coflow scheduling in classical data-parallel clusters\cite{chowdhury2015efficient}. 

    \item \textbf{Qubit capacity constraint:}  
     After reordering the arrived processes in queue by the priority, \tool{} greedily choose a batch from the process with the highest priority to the process with the lowest priority. The batch must fit on the device, also, the scheduler optimizes the hardware utility by the \emph{data occupancy ratio}. For example, if $\lambda=0.6$, \tool{} take a batch only when the sum of all data qubits uses at least 60\% of the hardware qubits.
    \[
     \lambda Q_{\mathrm{tot}}  \leq  \sum_{p \in B} q_p^d  <  Q_{\mathrm{tot}}.
    \]
    If all processes in the queue need less than $ \lambda Q_{\mathrm{tot}}$ data qubits, \tool{} keeps waiting until reach a maximum time bound that we set in the configuration file(10 second). 
    \item \textbf{Batch shot count:}
    For the selected batch $B$, \tool{} executes the minimum number of shots in the batch: 
    \[
        s_B = \min\!\bigl(S_{\max},\; \min_{p \in B} s_p\bigr),
    \]
    where $S_{\max}$ is a tunable maximum number of shots per batch.  
    This prevents very deep processes from creating long batches that waste shot resources, and allows
    deep jobs to be split across multiple batches. After the batch execution, each process's remaining shots is updated by subtracting $s_p\rightarrow s_p-s_B$. If $s_p=0$, the process finish and is removed from the queue. 
\end{enumerate}

\noindent\textbf{Example.}
Assume that the hardware device has $Q_{\text{tot}} = 24$ and we set $\lambda = 0.6$.
There should be at least $15$ data qubits to form a batch. Consider four processes waiting in the queue:
\begin{itemize}
\item $P_1$: $q^d_1 = 8$ qubits,\ $d_1 = 100$,\ $s_1 = 100$
\item $P_2$: $q^d_2 = 10$ qubits,\ $d_2 = 10$,\ $s_2 = 60$
\item $P_3$: $q^d_3 = 12$ qubits,\ $d_3 = 12$,\ $s_3 = 60$
\item $P_4$: $q^d_3 = 12$ qubits,\ $d_3 = 80$,\ $s_3 = 100$
\end{itemize}
A shot-and-depth unaware scheduler will take $P_1,P_2$ as the first batch and execute $60$ shots. This batch wastes space-time utility as $P_1$ and $P_2$ are misaligned in depth.
 \tool{} prioritizes $P_2,P_3$ first, since
$D_2 = 10 \cdot 60 = 600, D_3=12\times 60=720$ is the smallest two in \emph{total depth demand}.
The batch shot count is: $s_B = \min(S_{\max},\ \{s_2,s_3\}) = 60.$ Thus, \tool{} executes the batch $B$ containing $P_2,P_3$ for $60$ shots and updates process results accordingly.

\paragraph{Optimization objective.}
Suppose the scheduler forms a sequence of batches $B_1,\dots,B_K$.
For each batch $B_k$, let $D_{B_k} = s_k \cdot \max_{p \in B_k} d_p$
denote the batch makespan, where $s_k$ is the number of shots executed in batch $B_k$.
The corresponding space--time capacity is $\mathrm{Cap}_k = Q_{\mathrm{tot}} \cdot D_{B_k}$.
The useful work completed in batch $B_k$ is
$\mathrm{Work}_k = s_k \times \sum_{p \in B_k} q_p^d  d_p$. In general, we want to maximize the overall space-time utilization:
\begin{equation}
    \max_{\{B_k,s_k\}} 
\quad
\eta = 
\frac{\sum_{k=1}^{K} \mathrm{Work}_k}
     {\sum_{k=1}^{K} \mathrm{Cap}_k}
     \label{eq:STUtil}
\end{equation}
We calculate $\eta$ for the first batch in the previous example. If the first batch is $P_2,P_3$,  ${Work} = 60 \times (10\times 10+12\times 12)=14640$, and $\eta = \frac{Work}{Cap} = \frac{14640}{24 \times 60 \times 12}=0.85$. On the other hand if the first batch is $P_1,P_2$, ${Work} = 60 \times (8\times 100 +10\times 10)=54000$, and $\eta = \frac{Work}{Cap} = \frac{54000}{24 \times 60 \times 100}=0.375$. \tool{} improves $\eta$ by a factor of $2.27$.

\subsection{Data qubits space maneger}
\label{subsec:SpaceManager}

\begin{figure*}[ht!]
    \centering
    \includegraphics[width=0.8\linewidth]{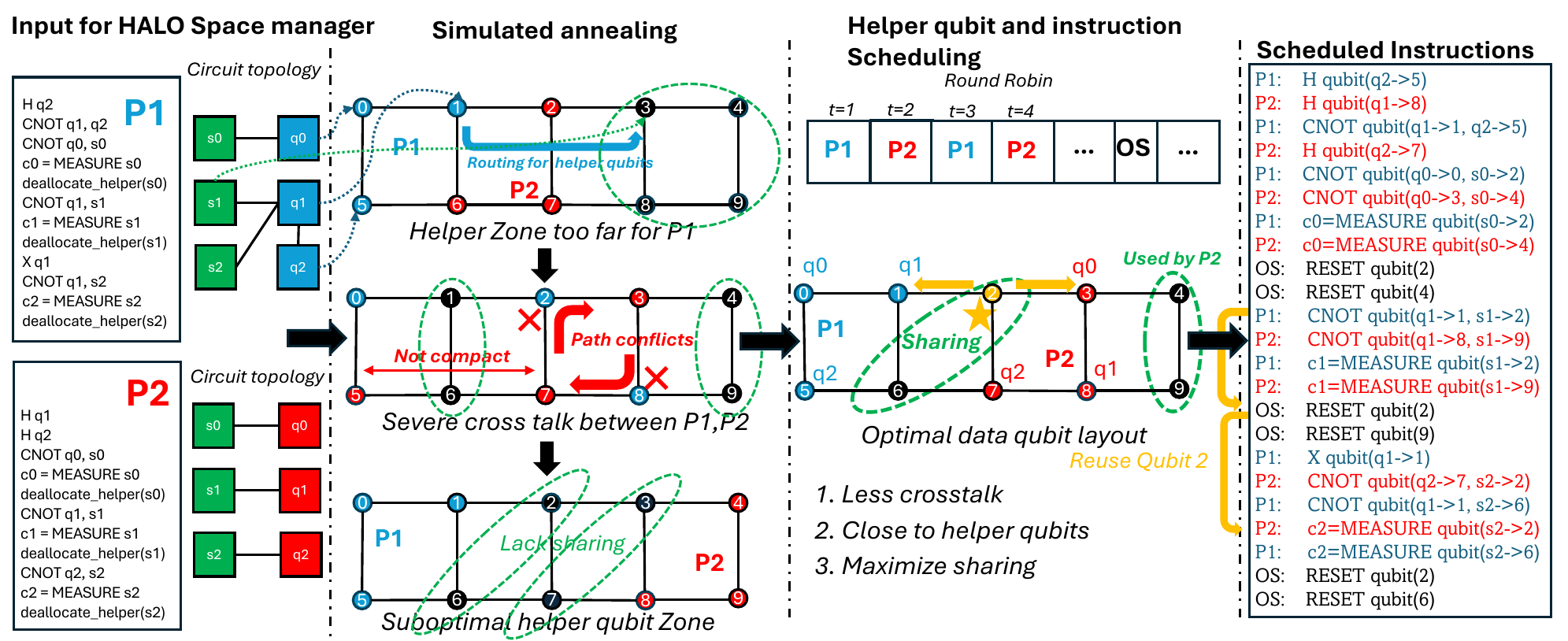}
    \caption{Example of HALO space management and instruction scheduling. \tool{} schedules a batch of two processes with $12$ qubits in total on a $10$ qubit quantum hardware(Impossible if not sharing helper qubits). Both $P1$ and $P_2$ have $3$ data qubits and $3$ helper qubits. We show three suboptimal data qubit layouts, and also the optimal data qubit layout.  Right: The full scheduled instructions of the two process batch in a round robin order after \tool{} fix the data qubit layout mapping.}
    \label{fig:placeholder}
\end{figure*}

Inspired by the quantum virtual machine \cite{tao2025quantum}, we further design a general cost function for our space-time scheduling algorithm. Our method has fewer constraints on hardware connections than HyperQ\cite{tan2020optimal}, where the hardware must have a strong space periodic structure. Despite the complexity of the problem, we capture three critical aspects that determine the mapping quality, and we formalize these aspects by cost functions. 

Our intuitions are as follows. First, data qubits cannot be shared and must stay in isolated territories, otherwise the computation of different processes might be entangled (\S \ref{sec:Background}). Also, data qubits of different processes should be as far from each other as possible to avoid cross-talk, thus meeting our design goal: \emph{isolation across processes} (\S \ref{sec:DesPrin}). Second, the helper qubit zone, which is defined as the rest of the qubits after mapping all data qubits, should be close to the data qubits in all processes that need helper qubits. This is our design goal for \emph{fine-grained resource sharing} (\S \ref{sec:DesPrin}). Ideally, all processes should get equal and fair access to the public helper qubit zone. Finally, the circuit topology of data qubits should match the layout, which is well studied in quantum circuit layout synthesis\cite{tan2020optimal}.

In \tool{}'s data qubit space manager, we configure the connection graph $G=(V,E)$ of a fixed hardware, where $V$ are hardware qubits and $E$ are hardware-qubit connections.

 Assume that the shot aware process batch scheduler (\S \ref{sec:shot-aware-schedul}) decides to execute $m$ processes $P_1,P_2,\cdots,P_m$. Each $P_i$ has $q^{d}_{P_i}$ data qubits in total and $q^{h}_{P_i}$ helper qubits. We denote the set of data qubits for process $P_i$ as $Q^d_i$. The data qubit manager in \tool{} further calculate a physical layout $L$ for these processes on $G$. The mapping $L$ balances routing efficiency among data qubits, crosstalk between multiple processes, and routing efficiency between data qubits and helper qubits. \tool{} iteratively optimizes a cost function that captures the quality of a joint placement across all processes. 
\paragraph{Inter-process crosstalk cost.}
 The first component of the cost function captures the level of mutual isolation between pairs of processes. This is defined as the average distance between all pair of data qubits $Q^d_a$ and $Q^d_b$ between two processes $P_a$ and $P_b$ in the mapping.  We denote this cost as $\text{Inter}(L,G)$.

\paragraph{Helper qubit routing cost.}
All the physical qubits that are not used as data qubits form the helper qubit zone $H$, where $H=V \setminus \bigcup_i \{L[q_i], q_i \in Q^d_i\}$. \tool{} has a greedy policy for the helper qubit mapping: any process will greedily find the nearest helper qubits when necessary. Thus, we measure the quality of data qubit layout by the extra routing cost between data qubits that need helper qubits with the helper qubit zone $H$. We call it \emph{helper qubit routing cost}, denoted as $\text{HRCost}(L,G)$.

\paragraph{Intro-process data qubit routing cost.}
The routing cost among data qubits is measured by the total number of extra number of SWAP gates needed with a fixed layout to connect remote data qubits, this is the same as the layout cost in the previous layout synthesis work\cite{tan2020optimal}.  We denote this cost as $\text{Intro}(L,G)$.

\paragraph{Total layout cost:}

Let $L:Q^d \to V$ be a layout mapping all virtual data qubits to hardware
qubits, and let $H \subseteq V$ be the helper region. For fixed non-negative
weights $\alpha,\beta,\gamma\in R_{\ge 0}$, the
total layout score is defined as:
\begin{equation}
    \begin{split}
                \text{Cost}(L,G)
    &=
    \alpha \,\text{Intro}(L,G)
    -
    \beta \,\text{Inter}(L,G)
    +\gamma \,\text{HRCost}(L,G)
    \end{split}
    \label{eq:LayoutCost}
\end{equation}
Our goal is to find a layout $L^\star$ that minimizes $ \text{Cost}(L,G)$. In this research, we set $\alpha=0.5,\beta=0.3,\gamma=1$.

In general, finding the optimal layout is NP-hard \cite{QubitMappingMaxSat}. We implemented a simulated annealing algorithm to iteratively search for a better mapping of data qubits onto a hardware graph. The algorithm begins with a greedy, process-clustered initial placement, then repeatedly proposes local moves, which are either small swaps within a process or moving a data qubit into a nearby helper site\cite{van1987simulated}.

\noindent\textbf{Example.}
In Figure \ref{fig:placeholder}, we show three different suboptimal layout mappings of data qubits for two processes P1,P2 during simulated annealing. 1. If \tool{} maps $\{5,0,1\}$ to P1 and $\{2,7,6\}$ to $P_2$, then $P_1$ is \emph{far from helping-zone}. Such a mapping increases the helper qubit mapping cost $\text{HRCost}(L,G)$.  If \tool{} map $\{0,2,8\}$ to P1 and $\{5,7,3\}$ to P2, then since there are routing path conflicts and the process mapping is not compact, the cross talk will severely affect process fidelity, this can be capture by a rather small inter-process cross talk cost $\text{Inter}(L,G)$. 3. If \tool{} map $\{0,1,5\}$ to P1 and $\{4,9,8\}$ to P2, P2 and P1 don't share helper qubits in between, the sharing opportunity is not optimized. \tool{} determines a much better data qubit layout mapping, where 
$\{0,1,5\}$ is mapped to P1 and $\{3,8,7\}$ is assigned to P2. In this layout, P1 and P2 can share hardware qubit $2$ in between, and the process cross talk is controlled.

\subsection{Helper qubit and instruction scheduler}
\label{subsec:HelperScheduler}

This section explains our design of the scheduling of process instruction with helper qubit dynamic mapping. Any virtual helper qubit $s_i$ is dynamically assigned a physical qubit during the instruction scheduling. Let $t$ bet the index of the current instruction finished being scheduled. $L^H(t)[s_i]=k$ means the virtual helper qubit $s_i$ at time $t$ is mapped to hardware qubit $k$.

In \tool{}, an instruction such as \texttt{CNOT q0, s0} is called a \emph{quantum virtual instruction}. The instruction scheduler in \tool{} replaces every virtual address with a real hardware address. The new translated instruction is called a \emph{quantum hardware instruction}. For example, if $L[q_0]=a,L^H[s_0]=b$, the translated hardware instruction of \texttt{CNOT q0, s0} is  \texttt{CNOT a, b} which is ready to be transpiled and executed.

In this stage, the output of the \tool{} instruction scheduler is a list of devirtualized \emph{quantum hardware instruction}. A \emph{Round-Robin} style scheduling algorithm is used in \tool{} to determine the instruction order. All processes have a fair chance to compete for the helper qubit resources when needed\cite{shreedhar1996efficient,rasmussen2008round,hahne2002round}. 

\tool{} schedules the next instruction for every process in the batch in turn. When process $P_a$ is selected to execute its next instruction $I$, \tool{} handles the instruction according to the following logic. If $I$ involves only data qubits, \tool{} directly translates it via the layout mapping $L$ and appends it to the scheduled instruction list. If $I$ requires helper qubit(s) and sufficient helper qubits are available, \tool{} greedily selects the nearest available helper qubit, allocates it to $I$, and schedules the translated instruction. Our design of  Helper qubit routing cost in \S \ref{subsec:SpaceManager} is consistent with this step, as the cost function measure how hard for a process to get nearby helper qubits when needed. If $I$ requires helper qubit(s) but none are available, \tool{} suspends $P_a$ and places it into a waiting state until helper qubits are released by other processes. Finally, if $I$ is a system call that releases a helper qubit, \tool{} inserts a hardware reset instruction and marks the corresponding qubit as available for future allocation. The reset instruction make sure that processes are not entangled (\S \ref{sec:Background}), and thus the \emph{process isolation} in design principle is satisfied (\S \ref{sec:DesPrin}).


\noindent\textbf{Example.}
The right part of Figure \ref{fig:placeholder} shows the full list of de-virtualized scheduled instruction of P1 and P2  take turns the run the next instruction. \tool{} inserts hardware reset instruction when helper qubits are released by system call. In the scheduled instructions, hardware qubit $2$ is first assigned to s1 in P1, then reset by \tool{}, and assigned to s2 in P2.  The example reveal the power of sharing helper qubits: P1 and P2 have $12$ qubits in total if we don't differentiate data qubits and helper qubits, and state of the art quantum space multiplexing system such as HyperQ can put at most one process in execution, with space utility at most $60\%$.

\section{Evaluation}
\label{sec:Eva}


To evaluate the effectiveness of \tool{} with respect to the challenges(\S ), we design experiments and answer the following research questions:
\begin{itemize}

\item \textbf{RQ1(\S \ref{subsec:RQ0})}: Can we tolerate the scheduling time overhead for a more fine grained hardware control in \tool{}?

\quad \emph{Yes}, we measure the scheduling time of \tool{} of some of the hardest cases in our benchmark suit, and the overhead is below $60$ seconds, which is mild compared with hours of waiting time.

\item \textbf{RQ2(\S \ref{subsec:RQ1})}: Does \tool{} perform better compared to state-of-the-art quantum runtime systems? \

\quad \emph{Yes}, we compare \tool{} against IBM Quantum and HyperQ to quantify improvements over existing cloud execution models. Specifically,\tool{} achieves 2.4$\times$ improvement in space utilization, a $4.44 \times$ compared with HyperQ.

\item \textbf{RQ3(\S \ref{subsec:RQ2})}: Is sharing helper-qubit really helpful for utilization and throughput? \;

\quad \emph{Yes}, we did ablation study by comparing \tool{} with \tool{}(No Sharing), which disables helper-qubit reuse to isolate the effect of spatial sharing. We observe that sharing helper qubits increase throughput by a factor $1.52 \times$ while increase the fidelity by $11\%$.

\item \textbf{RQ4(\S \ref{subsec:RQ3})}:Is space-time utility $\eta$ improved by shot-aware scheduling (\S \ref{sec:shot-aware-schedul})? \; \; 

\quad \emph{Yes}, we disable shot-adaptive scheduling to isolate temporal sharing benefits. The shot-aware batch scheduler in \tool{} consistently improves the space time utility $\eta$ by a factor of $2.87\sim 4.04$.
\item \textbf{RQ5(\S \ref{subsec:RQ4})}: Does throughput increase affect process fidelity after scheduling?  

\quad \emph{Yes}, we vary the data occupancy ratio $\lambda$(\S \ref{sec:shot-aware-schedul}) from $0.2$ to $0.8$ and observe that the average process fidelity in general decreases while the qubit sharing ratio and throughput increase. The primary contribution is extra noisy routing cost for helper qubits(HMCost in \S \ref{subsec:SpaceManager}) when throughput increase.  
\end{itemize}

\subsection{Setup}

\paragraph{Baselines.} We evaluate the performance of HALO scheduling against the following baselines.

\noindent\textbf{\textit{IBM Quantum}.}
The current approach used by the IBM Quantum Platform, and other quantum cloud providers, to run each program individually on the entire machine\cite{ibm-fairshare,ibm-busy}. 

\noindent\textbf{\textit{HyperQ}.}\;
The quantum virtual machine with fixed layout mapping pattern but not sharing helper qubits\cite{tao2025quantum}.

\noindent\textbf{\textit{HALO(NoSharing)}.}\;
\tool{} with shot scheduling and space multiplexing, but with helper-qubit sharing disabled. All qubits are seen as data qubits.

\noindent\textbf{\textit{HALO(ShotUnaware)}.}\;
\tool{} with helper-qubit sharing, but uses shot-unaware scheduling where the process batch a extract in an FIFO order.

\paragraph{Benchmarks.}
\begin{table}[t]
\centering
\renewcommand{\arraystretch}{1.00}
\small
\resizebox{\columnwidth}{!}{%
\begin{tabular}{c l l c c}
\hline
\textbf{Suite} & \# & \textbf{Name} & \textbf{Data} & \textbf{Helper} \\
\hline

\multirow{3}{*}{\textbf{I: QEC(Small)}} 
& 1 & shor\_stabilizer\_XZZX\_n3            & 3 & 5 \\
& 2 & syndrome\_extraction\_surface\_n4     & 4 & 4 \\
& 3 & cat\_state\_verification\_n4          & 4 & 2 \\

\hline
\multirow{3}{*}{\textbf{I: QEC(Medium)}} 
& 4 & fivequbit\_round\_2\_error\_IIZII            & 5 & 16 \\
& 5 & steane\_round\_2\_error\_IIIIIXI     & 7 & 12 \\
& 6 & surface\_d3\_round\_3\_error\_XIIIIIIII          & 9 & 24 \\
\hline

\multirow{3}{*}{\textbf{II: Arithmetic(Small)}} 
& 7 & varnum\_3\_d\_4\_1
& 3 & 19  \\
& 8 &
varnum\_3\_d\_4\_1
& 4 & 7 \\
& 9 &
varnum\_4\_d\_4\_1
& 4 & 8 \\

\hline
\multirow{3}{*}{\textbf{II: Arithmetic(Medium)}}

& 10 &
varnum\_10\_d\_3\_1
& 10 & 11 \\
& 11 &
varnum\_10\_d\_4\_0
& 10 & 19  \\
& 12 &
varnum\_12\_d\_4\_1
& 12 & 8  \\

\hline

\multirow{3}{*}{\textbf{III: Multi-CX(Small)}} 
& 13 & mcx\_2\_0   & 3 & 1 \\
& 14 & mcx\_3\_0   & 4 & 2 \\
& 15 & mcx\_4\_1   & 5 & 3 \\
\hline
\multirow{3}{*}{\textbf{III: Multi-CX(Medium)}} 
& 16 & mcx\_10\_1    & 11 & 9 \\
& 17 & mcx\_11\_0  & 12 & 10 \\
& 18 & mcx\_12\_0   & 13 & 11 \\
\hline
\multirow{6}{*}{\textbf{Random}} 
& 19 & data\_4\_syn\_4\_gc\_15\_0   & 4 & 4 \\
& 20 & data\_5\_syn\_5\_gc\_15\_0   & 5 & 5 \\
& 21 & data\_6\_syn\_6\_gc\_18\_0   & 6 & 6 \\
& 22 & data\_14\_syn\_15\_gc\_45\_0    & 14 & 15 \\
& 23 & data\_15\_syn\_12\_gc\_40\_0 & 15 & 12 \\
& 24 & data\_20\_syn\_10\_gc\_50\_0   & 20 & 10 \\
\hline

\end{tabular}
}
\caption{Some examples in our benchmark suites, which consist of: (I) helper-heavy QEC stabilizer circuits;  (II) arithmetic circuits of varying sizes; 
(III) multi-controlled $X$ circuits; Below are examples of some randomly generated circuits.
}
\label{tab:unified_benchmarks}
\end{table}
Currently, existing quantum circuit benchmarks such as QASMbench\cite{li2023qasmbench}, do not distinguish between data qubits and helper qubits. To address this limitation, we introduce a new benchmark suite specifically designed to capture this distinction. Our suite consists of four categories of benchmarks with varying circuit sizes and helper-qubit proportions, which we believe are representative of near-term quantum applications that rely on helper qubits as well as the future fault-tolerant quantum computers: (I) Stabilizer measurement circuits(QEC) where all helper qubits served to detect errors\cite{shor1996fault,riste2015detecting}. We injected a single artificial Pauli error in some programs.  (II) Arithmetic circuits. We generate a lot of boolean arithmetic circuits with varying structures and number of variables~\cite{thomsen2010reversible,haener2018quantum}. (III) A set of Multi-controlled $X$ circuits($C^n(X)$), or quantum condition circuit, which is the most important building block for universal quantum programs and is hard to compile without helper qubits\cite{conditional,zindorf2025efficientimplementationmulticontrolledquantum,miller2011elementary}. Each benchmark has both small scale, and medium scale. (IV) A mix benchmark, which includes all the above with extra randomly generated circuits. Some of the concrete circuit examples are listed in Table~\ref{tab:unified_benchmarks}.

\paragraph{Metric.} We evaluate the performance of \tool{} and baselines based on the following metrics:

\noindent\textbf{\textit{Space Utilization}}
Space utilization quantifies how effectively the hardware’s physical qubits are used during execution. It measures the fraction of qubits allocated to active processes(consider both helpers qubit and data qubits) relative to the total number of qubits on the hardware device.

\noindent\textbf{\textit{Share Ratio}.}
Share Ratio measures how effectively \tool{} enables helper-qubit reuse across concurrent processes. It is defined as the fraction of helper qubits that are used by at least two processes during execution. Let $H$ denote the set of all helper qubits, and let $\text{count}(q)$ be the number of processes that use helper qubit $q$. The Share Ratio is computed as:
\begin{equation}
    \text{ShareRatio}
    = \frac{\left| \{\, q \in H \mid \text{count}(q) \ge 2 \,\} \right|}
           {|H|}.
           \label{eq:ShareRatio}
\end{equation}
In helper qubit intense application, a higher Share Ratio indicates that more helper qubits are successfully reused, leading to improved spatial utilization and increased parallelism.

\noindent\textbf{\textit{Space-Time Efficiency $\eta$.}}
We use the average space-time efficiency $\eta$ per batch to evaluate the effectiveness of our shot-aware batch scheduling (\S \ref{sec:shot-aware-schedul}). A higher $\eta$ indicates that the hardware is performing proportionally more useful work rather than remaining idle due to batching inefficiencies.

\noindent\textbf{\textit{Process per Batch (PPB)}.} We define \emph{processes per batch} (PPB) as the average number of concurrently executing quantum processes within each scheduled batch. A higher value means larger throughput and lower latency. 

\noindent\textbf{\textit{Fidelity.}}
We compare the measured output distribution of each circuit against its ideal (noise-free) distribution obtained from a state-vector simulator. We adopt the $L_1$ distance as our fidelity metric.

\begin{equation}
    L_1 = \sum_{s \in \mathcal{S}} \left| p_{\text{ideal}}(s) - p_{\text{real}}(s) \right|,
\end{equation}
where $\mathcal{S}$ is the set of all possible measurement outcomes. The reported fidelity is computed as $1 - \frac{1}{2}L_1$, so that higher values indicate closer agreement with the ideal distribution.

\paragraph{Experimental Setup.}
To evaluate the performance of the \tool{} scheduling framework under realistic job submission patterns, 
we conducted experiments using the standard Poisson distribution as used in HyperQ, where jobs arrive following a Poisson process with a mean rate of $0.6$ job per second. We keep generate random job for $40$ seconds in all experiments. After scheduling by \tool{} or other baseline method, we transpiled the circuit and submit through IBM cloud and wait for the result. We did separate experiments to study the space scheduling and the shot-aware time scheduling. In space scheduling experiment, we fix all shots to be the $1000$. In the experiment of shot-aware time scheduling, all processes have a randomly generated shots ranges between $500$ to $8000$. 

\begin{figure}[t!]
    \centering
    \setlength{\fboxsep}{2pt}

    \fbox{\includegraphics[width=0.2\textwidth]{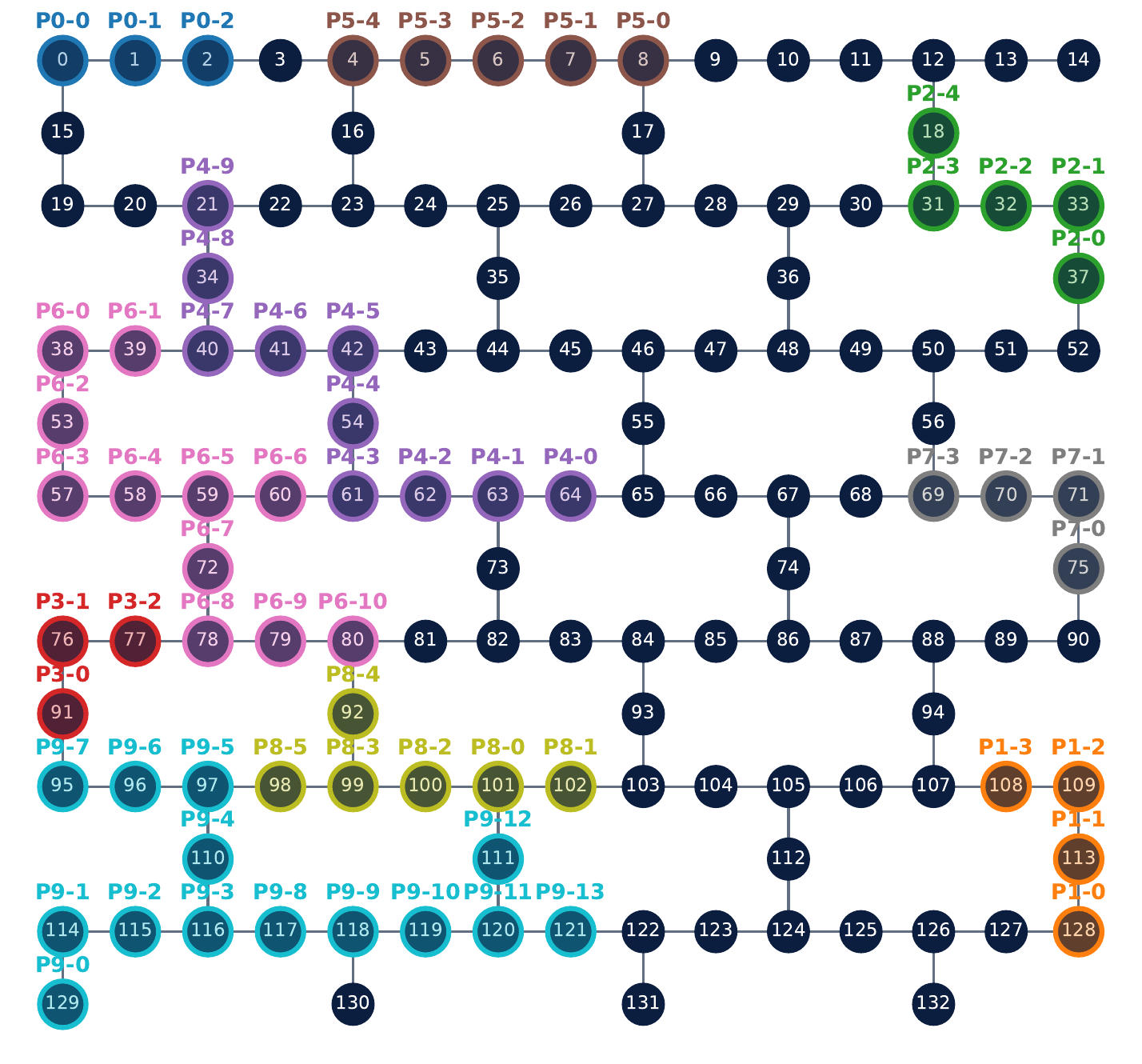}}
    \hspace{0.3cm}
    \fbox{\includegraphics[width=0.2\textwidth]{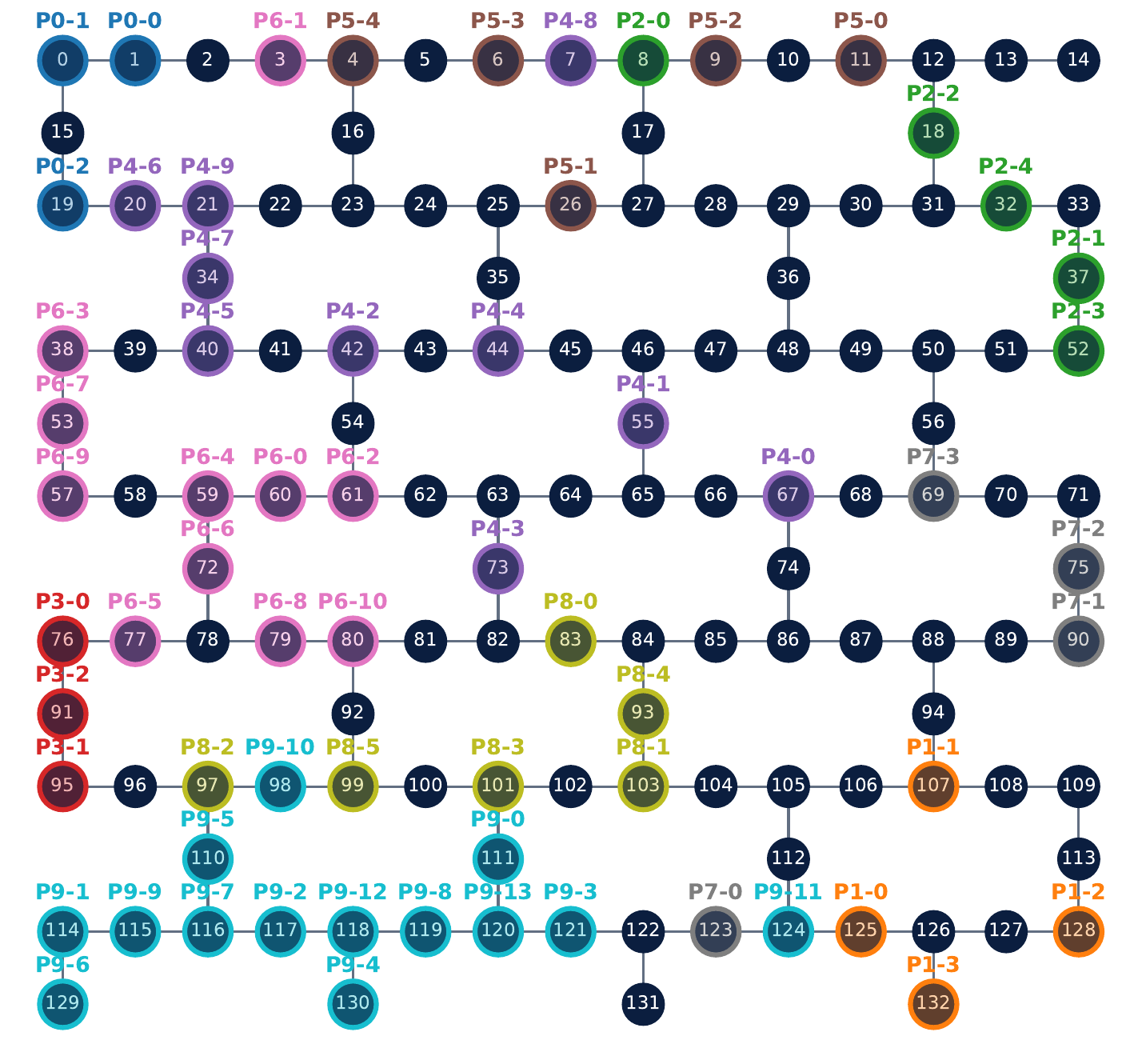}}

    \caption{Left: The greedy cluster initial mapping of \tool{} for a batch with 10 processes on IBM Torino quantum processor.  
    Right: Final mapping decision of \tool{} after 30 iterations of simulated annealing.}
    \label{fig:exampleHALO}
\end{figure}

All experiments were executed on the \emph{IBM Quantum Platform} using the \emph{IBM Torino} quantum processor, 
which features a \emph{133-qubit square lattice} architecture available for public access. Figure \ref{fig:exampleHALO} shows a concrete example of \tool{} manage data qubits of 10 processes on the hardware. Compilation, scheduling, and aggregation of logical jobs into larger composite circuits were performed on a 
\emph{desktop workstation} equipped with an 12th Gen Intel(R) Core(TM) i5-12400K CPU (2.5\,GHz) and 64\,GB RAM. All scheduled circuits are further transpiled to the targeted hardware using \emph{qiskit v2.2.3} with optimization level $3$.

\subsection{RQ1: Scheduling overhead}
\label{subsec:RQ0}

\begin{figure}[h!]
    \centering
    \includegraphics[width=0.33\textwidth]{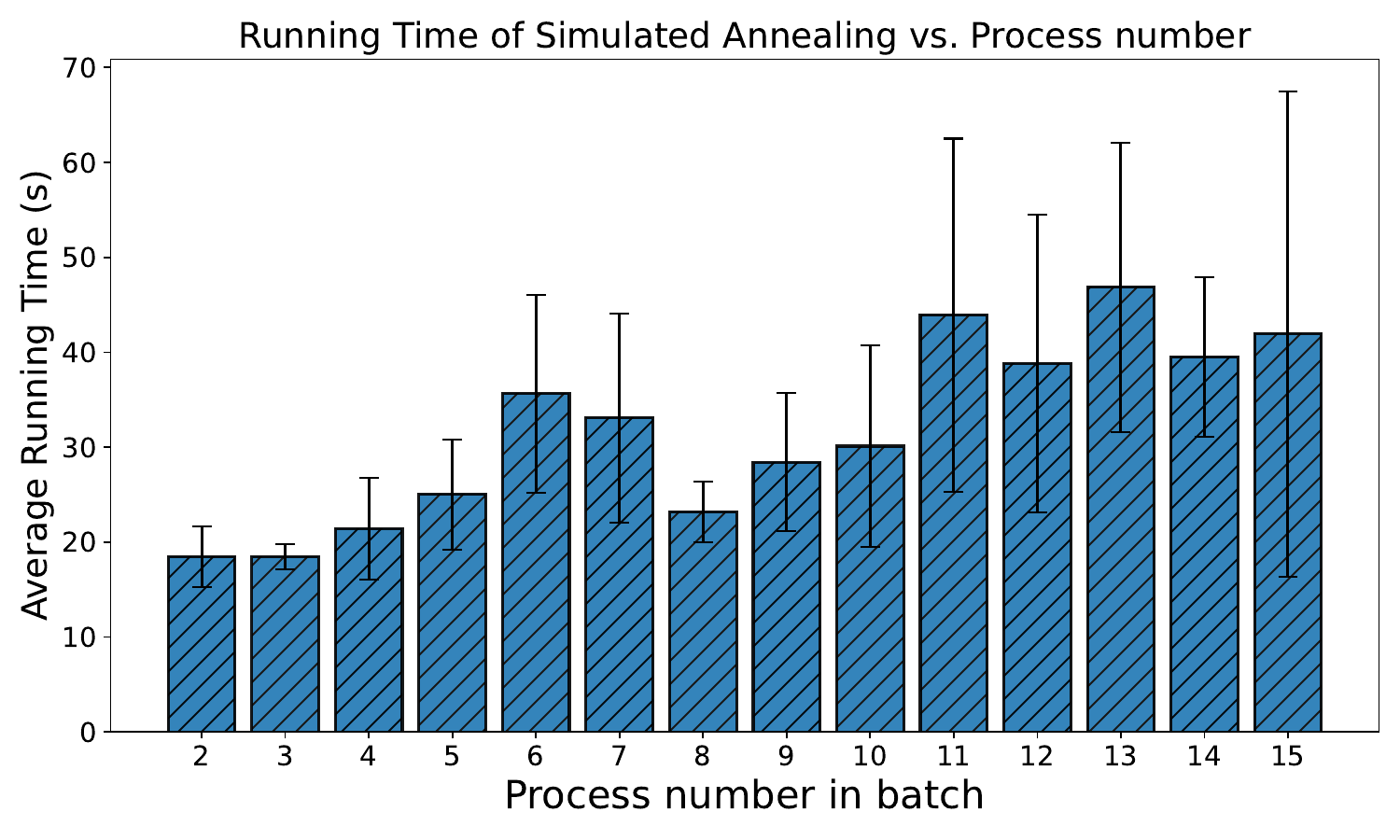}
    \caption{The average running time of \tool{}'s space management with different number of processes in batch(\S \ref{subsec:SpaceManager}). The scheduling ends within $60$ seconds for almost all cases, much less than hours of waiting time in current quantum cloud service.}
    \label{fig:timeDynamic}
\end{figure}

In experiments, we observe that the running time of shot-aware batch scheduling (\S \ref{sec:shot-aware-schedul}) and instruction scheduling (\S \ref{subsec:HelperScheduler}) with fixed data qubit layout mapping is negligible (<$0.1$ seconds) compared with the space layout optimization (\S \ref{subsec:SpaceManager}), which is the real possible bottleneck. We stress test the total running time of \tool{} space scheduling of \tool{} with a fixed number of processes in one batch ranges from $2$ to $15$. As shown in Figure~\ref{fig:timeDynamic}, space scheduling typically completes within one minute, which is negligible relative to the hours of queueing delay on IBM Quantum~\cite{tao2025quantum,ibm-busy}. Thus, we conclude that \tool{}'s scheduling overhead is acceptable in the near-term devices with hundreds of qubits. 

\begin{figure*}[tbp]
    \centering
    \includegraphics[width=0.8\textwidth]{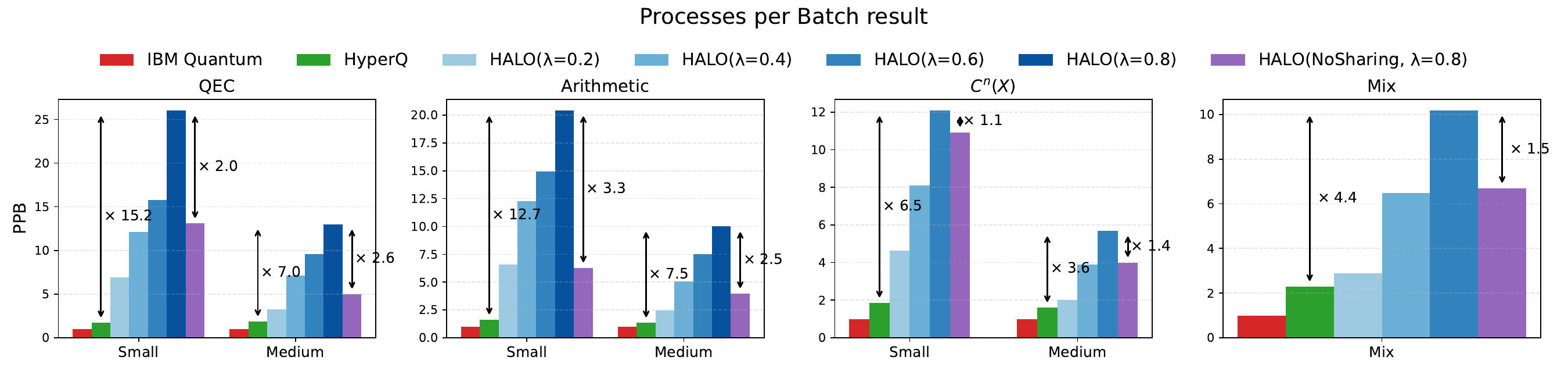} 
    \includegraphics[width=0.8\textwidth]{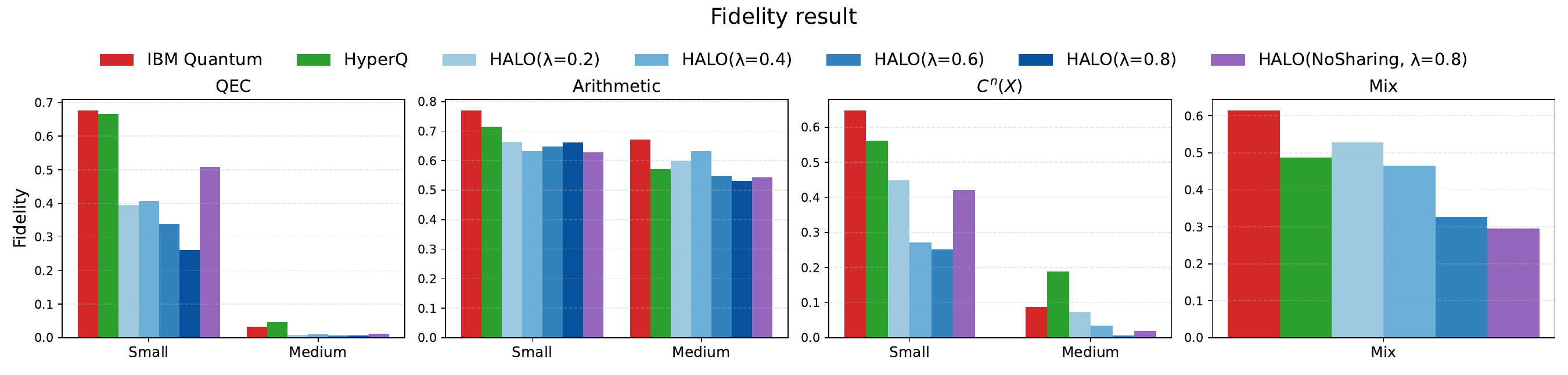}   
    \caption{Result of our evaluation. Above: Result of process per batch (PPB) on three benchmarks and the mix benchmark. \tool{} improve PPB by a factor of $\times 15.2,\times 12.7$ compared with HyperQ at when $\lambda=0.8$ and the utilization is maximized in the first two benchmarks, and a factor $\times 6.5, \times 4.4$ when $\lambda=0.6$ in the last two benchmarks. Below: Result of average fidelity. \tool{} remains competitive fidelity in Arithmetic benchmark and Mix benchmark, while is more sensitive to noise in QEC and $C^n(X)$ benchmark.}
    \label{fig:result}
    \vspace{-2ex}
\end{figure*}

\begin{table}[t]
\centering
\renewcommand{\arraystretch}{1.1}
\resizebox{\columnwidth}{!}{%
\begin{tabular}{l l c c c c c c}
\hline
 & 
& \multicolumn{2}{c}{\textbf{Space Utilization}}
& \multicolumn{2}{c}{\textbf{Fidelity}}
& \multicolumn{2}{c}{\textbf{PPB}} \\
\textbf{Benchmark} & \textbf{Algorithm} 
& \textbf{Small} & \textbf{Medium}
& \textbf{Small} & \textbf{Medium}
& \textbf{Small} & \textbf{Medium} \\
\hline

\multirow{4}{*}{QEC}
& IBM Quantum  & 0.067 & 0.187 & 0.676   & 0.033  & 1 & 1 \\
& HyperQ       & 0.111 & 0.353 & 0.666   & 0.046  & 1.71 & 1.85 \\
& HALO (0.6)   & 0.618 & 0.669 & 0.340 & 0.007 & 15.79 & 9.58 \\
\hline

\multirow{4}{*}{Arithmetic}
& IBM Quantum  & 0.136 & 0.242 & 0.769   & 0.671   & 1 & 1 \\
& HyperQ       & 0.292 & 0.309 & 0.714   & 0.571   & 1.6 & 1.33 \\
& HALO (0.6)   & 0.817 & 0.810 & 0.647 & 0.547 & 14.92 & 7.54 \\
\hline

\multirow{4}{*}{$C^n(X)$}
& IBM Quantum  & 0.067 & 0.176 & 0.648  & 0.033  &  1 & 1 \\
& HyperQ       & 0.128 & 0.279 & 0.666   & 0.046   & 1.85 & 1.6 \\
& HALO (0.6)   & 0.739 & 0.923 & 0.340 & 0.007 & 12.08 & 5.69 \\
\hline

\multirow{4}{*}{Mix}
& IBM Quantum  & \multicolumn{2}{c}{0.133} & \multicolumn{2}{c}{0.614} & \multicolumn{2}{c}{1} \\
& HyperQ       & \multicolumn{2}{c}{0.330} & \multicolumn{2}{c}{0.487} & \multicolumn{2}{c}{2.29} \\
& HALO (0.6)   & \multicolumn{2}{c}{0.805\textcolor{red}{($\uparrow$\,2.44$\times$)}} & \multicolumn{2}{c}{0.327\textcolor{blue}{($\downarrow$\,33\%)}} & \multicolumn{2}{c}{10.17 \textcolor{red}{($\uparrow$\,4.44$\times$)}} \\
\hline

\end{tabular}
}
\caption{Space utilization (utility), fidelity, and processes-per-batch (PPB) across benchmarks and scheduling configurations for \tool{} ($\lambda=0.6$) vs. IBM Quantum vs. HyperQ. In the Mix benchmark, \tool{} achieves 2.44$\times$ improvement in space utilization, a $4.44 \times$ improvement in throughput while keeps the fidelity loss within $33\%$ compared with HyperQ.}
\vspace{-2ex}
\label{tab:mainresult}
\end{table}

\subsection{RQ2: Overall performance}
\label{subsec:RQ1}
We compare \tool{} (fixing the data-occupancy ratio at $\lambda=0.6$) against IBM Quantum (exclusive-use scheduling) and HyperQ (logical multiplexing without hardware sharing)\cite{tao2025quantum}. As shown in Figure~\ref{fig:result} and Table \ref{tab:mainresult}, \tool{} substantially increases the resource utilization and processes per batch relative to both baselines. Compared to the IBM Quantum execution model, which serializes all jobs, \tool{} increases resource utilization by 5.34 $\times$ on average in all benchmarks, while increasing the processes per batch by 10.17 $\times$ on average, which effectively saves $90.1\%$ of end-to-end latency for users in a throughput-limited regime. Compared to HyperQ, \tool{} achieves a 2.99 $\times$ improvement in utilization and a 4.44 $\times$ increase in processes per batch by jointly removing HyperQ's fixed virtual-machine layout constraint and enabling explicit helper-qubit sharing, both of which are unavailable in HyperQ and jointly responsible for its severe fragmentation.

For fidelity, IBM Quantum and HyperQ maintain similar performance (49\% and 46\% on average), while \tool{} introduces a fidelity reduction of approximately 35\% due to increased crosstalk and extra routing of qubit sharing (\S \ref{subsec:SpaceManager}). However, this degradation is a tunable consequence of aggressive resource sharing and can be significantly reduced by lowering the data-occupancy ratio $\lambda$ (discuss in \S ~\ref{subsec:RQ4}).

Take Benchmark II (the arithmetic benchmark, small circuits) as an example, where we populate the waiting queue with a mix of jobs that vary in both qubit usage. Under IBM Quantum, the system executes jobs strictly in FIFO order after another, causing workloads with small space usage to dominate the hardware and resulting in low space utility (Table~\ref{tab:mainresult}): IBM Quantum reaches a utility of 0.136, fidelity of 0.769, and only 1 process per batch. HyperQ mitigates this issue by enabling logical co-scheduling, increasing utility to 0.292 and processes per batch to 1.6, with fidelity 0.714. 

In contrast, with $\lambda=0.6$, \tool{} concurrently executes many more arithmetic jobs by reusing shared helper qubits and grouping tasks according to their space demands. On the same benchmark, \tool{} achieves utility 0.817 and 14.9 processes per batch, i.e., about a 9.3$\times$ throughput improvement over HyperQ (and 15$\times$ over IBM Quantum), while keeping fidelity at 0.647, only modestly below HyperQ (0.714). This illustrates how helper-qubit sharing delivers order-of-magnitude throughput gains at the cost of a relatively small fidelity reduction on practical workloads.

\subsection{RQ3: Effectiveness of qubit sharing}
\label{subsec:RQ2}

\begin{table}[t]
\centering
\renewcommand{\arraystretch}{1.15}
\resizebox{\columnwidth}{!}{%
\begin{tabular}{l l c c c c}
\hline
\textbf{Benchmark}
& \textbf{Configuration}
& \multicolumn{2}{c}{\textbf{Fidelity}}
& \multicolumn{2}{c}{\textbf{PPB}} \\
\cline{3-4} \cline{5-6}
& & Small & Medium & Small & Medium \\
\hline

\multirow{2}{*}{QEC}
& HALO (0.6)            & 0.262 & 0.007 & 26.00 & 13.00 \\
& HALO (No Sharing)    & 0.508 & 0.012 & 13.08 & 5.00  \\
\hline

\multirow{2}{*}{Arithmetic}
& HALO (0.8)            & 0.661 & 0.532 & 20.40 & 10.00 \\
& HALO (No Sharing)    & 0.628 & 0.544 & 6.27  & 3.96  \\
\hline

\multirow{2}{*}{$C^n(X)$}
& HALO (0.6)            & 0.251 & 0.007 & 12.08 & 5.69  \\
& HALO (No Sharing)    & 0.420 & 0.020 & 10.93 & 4.00  \\
\hline

\multirow{2}{*}{Mix}
& HALO (0.6)            & \multicolumn{2}{c}{0.327 \textcolor{red}{($\uparrow$\,11\%)}} & \multicolumn{2}{c}{10.17 \textcolor{red}{($\uparrow 1.52\times$ )}} \\
& HALO (No Sharing)    & \multicolumn{2}{c}{0.295 } & \multicolumn{2}{c}{6.69}  \\
\hline
\end{tabular}
}
\caption{Comparison of fidelity and processes per batch (PPB) between HALO and HALO (No Sharing, but maximize space multiplexing).In the Mix benchmark, sharing helper qubits increase throughput by a factor $1.52 \times$ while increase the fidelity by $11\%$. }
\label{tab:halo-vs-noshare-fidelity-ppb}
\end{table}

\begin{figure}[htbp]
    \centering
    \includegraphics[width=0.47\textwidth]{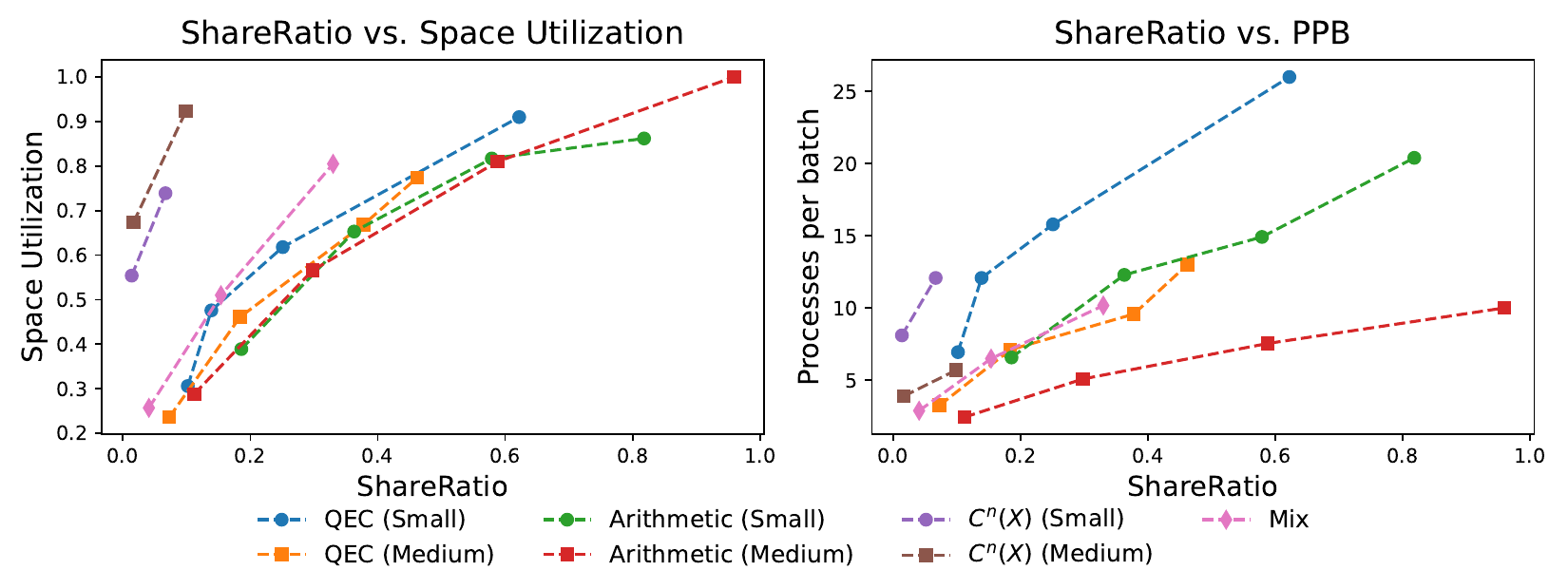}
    \caption{We plot the shareRatio with space utilization and PPB for all our benchmark. The result consistently show that the more qubit we allow \tool{} to share, the larger the throughput and the space utilization.}
    \label{fig:SHare}
    \vspace{-2ex}
\end{figure}

To isolate the contribution of qubit resource sharing, we perform an ablation study comparing \tool{} to  \tool{}(No Sharing), which disables helper-qubit reuse while retaining the same placement and scheduling strategy. As shown in Figure~\ref{fig:result} and Table~\ref{tab:halo-vs-noshare-fidelity-ppb}, \tool{} substantially increases processes per batch by allowing concurrently executing jobs to share helper qubits rather than reserving private helper qubit regions.Without sharing, a large fraction of physical qubits are forced to remain idle due to exclusive helper qubit reservation, particularly for workloads with high ancilla-to-data ratios such as the \emph{Arithmetic} benchmark, where reversible logic synthesis introduces a heavy helper-qubit footprint\cite{thomsen2010reversible,haener2018quantum}. In this setting, HALO(No Sharing) achieves only 6.27 and 3.96 processes per batch (small/medium), whereas \tool{} reaches 20.40 and 10.00, corresponding to a $3.25\times$ and $2.53\times$ improvement in throughput, respectively. This translates to a 69--76\% reduction in average waiting time for users under sustained load. In the Mix benchmark, \tool{}($0.6$) increase throughput by a factor of $1.52 \times $ while increase the fidelity by $11\%$. 

Figure \ref{fig:SHare} indicates that the increase in space utility and throughput is a direct result of sharing helper qubits. The utility and PPB increase consistently with ShareRatio across all benchmarks. In the most aggressive case, Share ratio approaches $1$, which indicates that almost all helper qubits are used by more than $2$ different processes. The results match our expectation illustrated in Figure \ref{fig:share}(\S \ref{sec:Intro}). In conclusion, we prove the value of sharing helper qubits: \emph{The more we share helper qubits, the larger throughput gain we will have}.




\subsection{RQ4: Effectiveness of shot-aware batch scheduling}
\label{subsec:RQ3}

\begin{table}[th!]
\centering
\renewcommand{\arraystretch}{1.0}
\resizebox{\columnwidth}{!}{%
\begin{tabular}{l c c c c c}
\hline
& \multicolumn{3}{c}{\textbf{Space time utility $\eta$}}
& \multicolumn{2}{c}{\textbf{Fidelity}} \\
\cline{2-4} \cline{5-6}
& Aware & Unaware & Improve & Aware & Unaware \\
\hline
HALO(0.2) &  0.235    &  0.082    & $\times 2.87$     &   0.493    &   0.483   \\
HALO(0.4) &  0.373    &  0.083    &  $\times 4.49$   &  0.427     &  0.429     \\
HALO(0.6) &  0.460    &  0.114    &   $\times 4.04$  &   0.380    &  0.415    \\
\hline
\end{tabular}%
}
\caption{Space--time utilization and fidelity on the Mix benchmark
for shot-aware and shot-unaware scheduling.}
\label{tab:mix-shotaware-vs-unaware}
\end{table}

 To study the effectiveness of our Shot-aware batch scheduling algorithm, we run HALO scheduling on a process queue randomly selected from all benchmark circuits. All process has a randomly generated shots value that ranges between $500$ to $8000$.  We run the same experiments for HALO algorithm with both 1. \emph{Shot-unaware} batch scheduling, where the next batch is taken from the queue according to the FIFO order. 2. \emph{Shot-aware} batch scheduling (\S \ref{sec:shot-aware-schedul}). We tune the \emph{data occupancy ratio} $\lambda$ from $0.2$ to $0.6$ and repeat the same experiment. The results in Table \ref{tab:mix-shotaware-vs-unaware} shows our Shot-aware batch scheduling consistently improve the space time utility $\eta$ by a factor of $\times 2.87,\times 4.49,\times 4.04$. Moreover, the fidelity of \tool{} with shot-unaware batch scheduling is close to the fidelity with shot-aware batch scheduling. We conclude that shot-aware scheduling does not harm the fidelity.


\subsection{RQ5: Throughput-fidelity trade-off and explanation}
\label{subsec:RQ4}

We observe the tradeoff between the averaged fidelity and the throughput. When aggressive qubit sharing is allowed, the fidelity of all benchmarks degrade. Especially for the QEC benchmark and $C^n(X)$ benchmark. 

\begin{figure}[t]
    \centering

    \begin{subfigure}[t]{0.48\columnwidth}
        \centering
        \includegraphics[width=\linewidth]{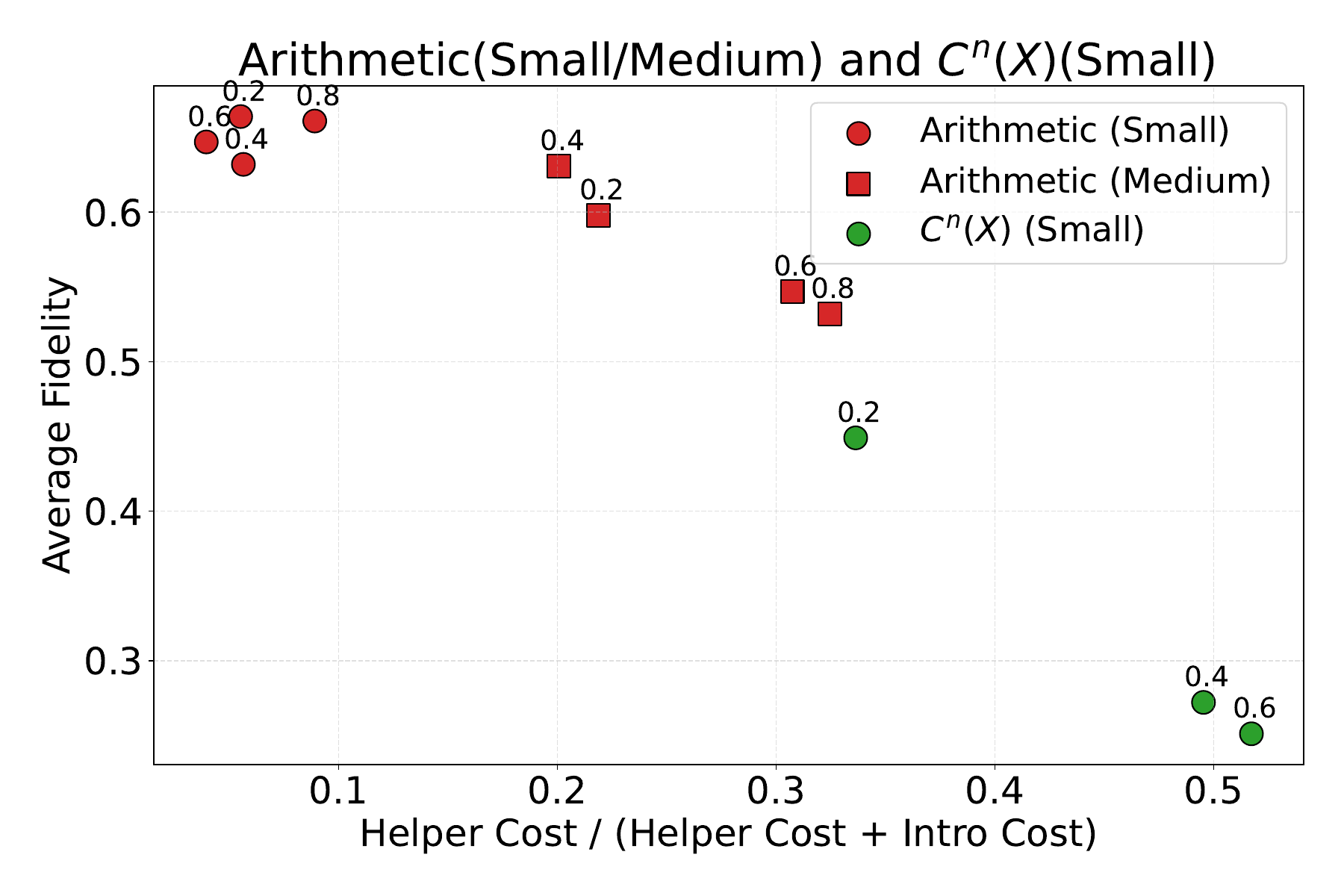}
        \caption{}
        \label{fig:costRatio}
    \end{subfigure}
    \hfill
    \begin{subfigure}[t]{0.48\columnwidth}
        \centering
        \includegraphics[width=\linewidth]{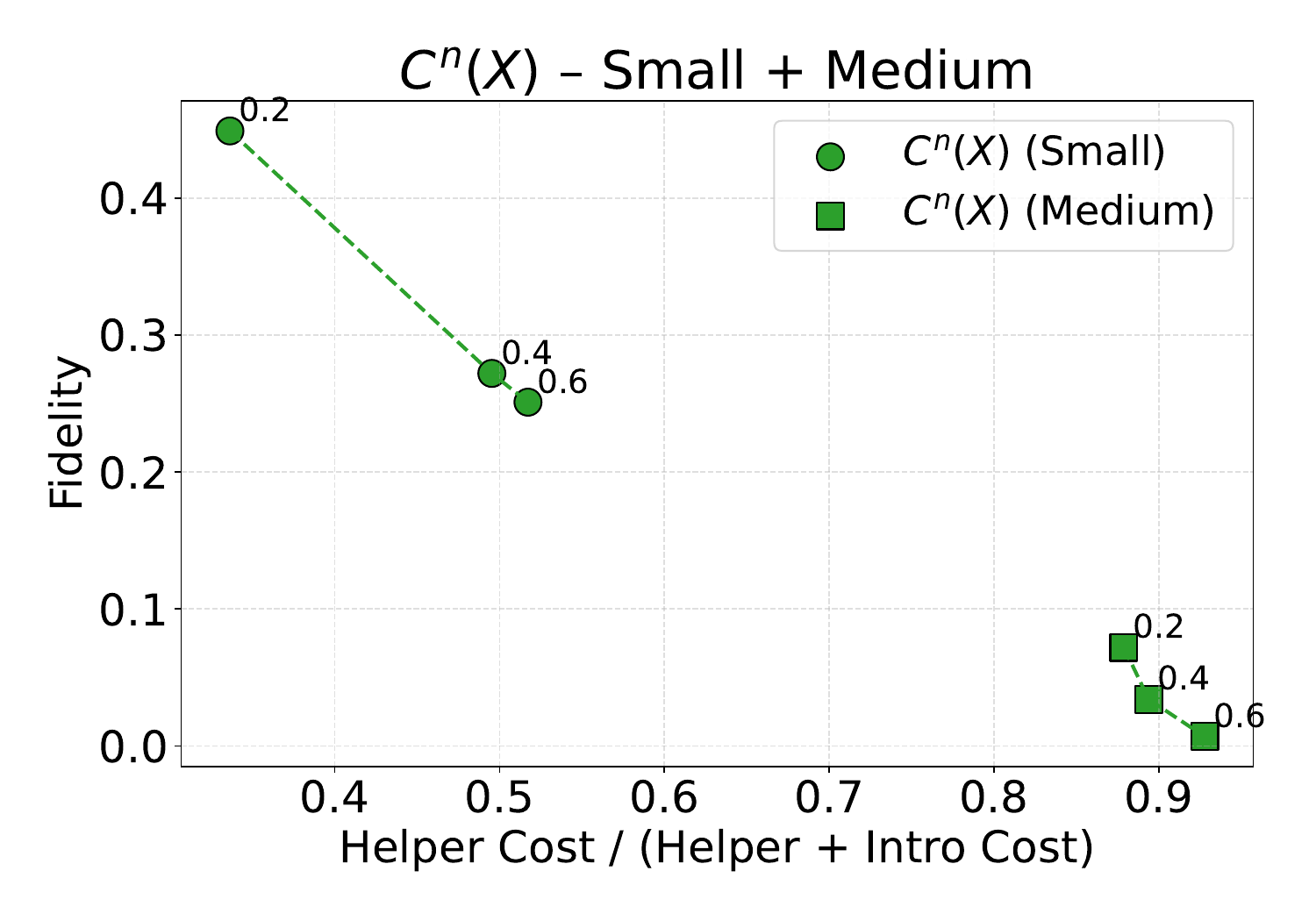}
        \caption{}
        \label{fig:cnx}
    \end{subfigure}

    \vspace{0.6em}

    \begin{subfigure}[t]{0.9\columnwidth}
        \centering
        \includegraphics[width=\linewidth]{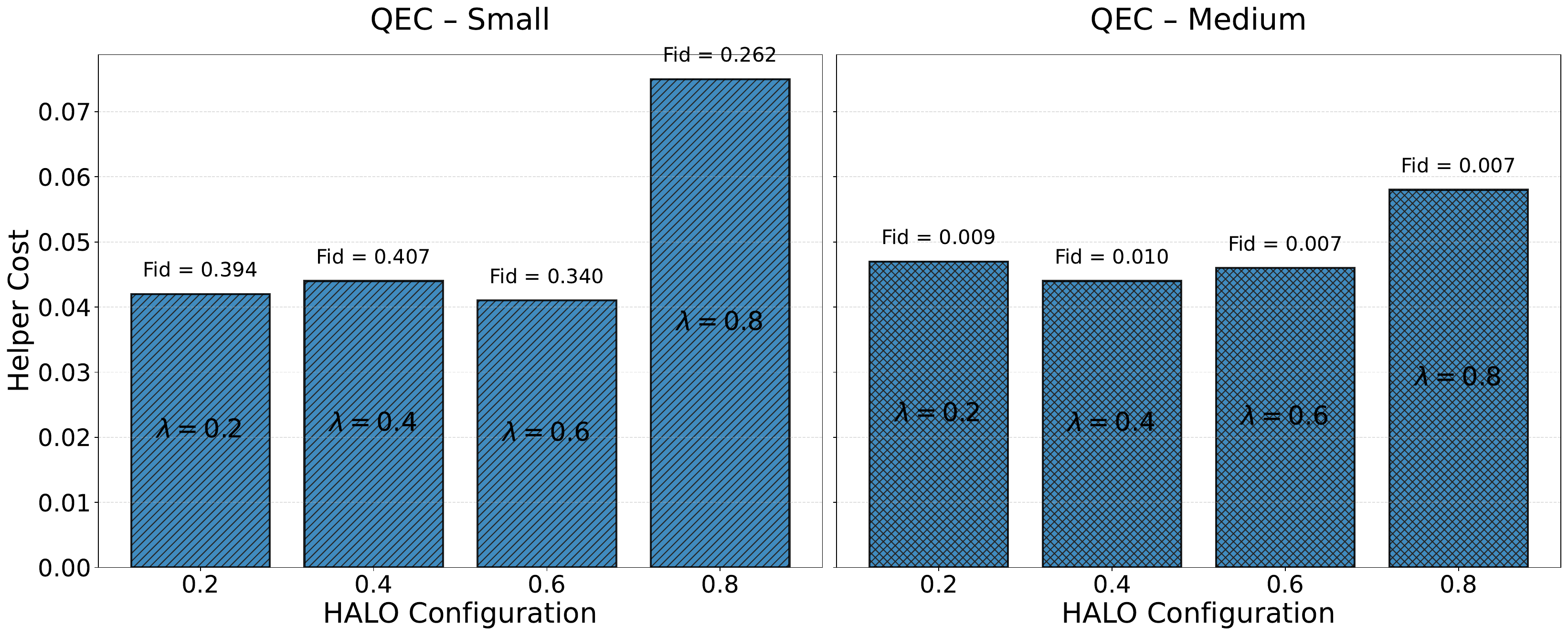}
        \caption{}
        \label{fig:qec}
    \end{subfigure}

    \caption{We study the relation between fidelity with $\text{HRCost}(L,G)$ and ratio of helper qubit routing cost in \tool{}:
    $\frac{\text{HRCost}(L,G)}{\text{Intro}(G,L)+\text{HRCost}(L,G)}$. In (a),(b), we observe a strong correlation between the ratio with the final fidelity. In QEC benchmark, the ratio is always $1$, and the fidelity degration can be explained by increase of $\text{HRCost}(L,G)$}
    \vspace{-2ex}
    \label{fig:threeInTwoRows}
\end{figure}

To understand the difference between the fidelity degradation level in different benchmarks, we analyze all the cost functions of the results(\S \ref{subsec:SpaceManager}). We can explain the difference well by the ratio of routing cost of helper qubit, calculated as $\frac{\text{HRCost}(L,G)}{\text{Intro}(G,L)+\text{HRCost}(L,G)}$(\S \ref{subsec:SpaceManager}). which quantifies how much of the total routing overhead comes specifically from accessing helper qubits. A higher ratio indicates that a larger fraction of noise is
introduced by helper-qubit routing and reuse. This ratio is highly sensitive to the $\lambda$, as $\lambda$ increases, more physical qubits are allocated as data qubits, forcing helper qubits to be shared across more processes. This increases both the distance and frequency of data–helper interactions, and thus the helper-routing cost. In the QEC benchmark, where each data qubit depends on many helper qubits, the ratio is always~1, meaning all additional routing noise is caused by accessing helper qubits—this explains why \tool{} performs worst on QEC workloads.

In contrast, for Arithmetic and $C^n(X)$ circuits, the fidelity decreases roughly linearly with this ratio (Figure~\ref{fig:threeInTwoRows}). The medium-scale $C^n(X)$ benchmark represents the worst non-QEC case: almost all routing overhead comes from helper access, resulting in fidelity below 10\%. Conversely, in the Arithmetic benchmark, less than half of the routing cost originates from helper-qubit
interaction, allowing \tool{} to maintain substantially higher fidelity even as
$\lambda$ increases.

For QEC circuits which are are heavily ancilla-intensive and $\text{Intro}(G,L)$ is always $0$, we study the fidelity degradation with $\text{HRCost}(L,G)$. Bar plot (c) in Figure \ref{fig:threeInTwoRows} indicates the degradation comes from the increase of $\text{HRCost}(L,G)$ cost itself. In summary, helper qubit routing and sharing have a more severe effect on circuit fidelity than intro circuit data qubit routing in \tool{}. Despite the tradeoff, we expect that \tool{}'s performance will keep increasing with the improvement of quantum hardware towards fault-tolerance in the next few years\cite{evered2023high,google2023suppressing}.

\section{Related Work}


\paragraph{Quantum virtualization and multi-tenant execution.}
Several recent systems explore ways to allow multiple quantum programs to coexist on a single device. HyperQ~\cite{tao2025quantum} introduces virtual quantum machines (qVMs) to multiplex circuits on cloud hardware and demonstrates that concurrency can reduce user-perceived latency. QOS~\cite{giortamis2024qos} similarly provides abstractions for job submission, device sharing, and execution monitoring. While these systems significantly advance the software stack for cloud quantum computing, they do not exploit opportunities for fine-grained physical resource sharing. In contrast, \tool{} enables both spatial and temporal sharing at the hardware level by reusing helper qubits and incorporating shot-adaptive scheduling.

\paragraph{Resource sharing in classical operating systems.}
Resource sharing is a foundational capability of classical operating systems, which multiplex CPU, memory, and I/O using long-established mechanisms such as time-sharing, virtual memory~\cite{Goldberg1974VirtualMachines}, and modern isolation tools such as cgroups~\cite{menage2007cgroups}. User-level threading and scheduler-activation techniques further illustrate how CPU and execution contexts can be multiplexed efficiently~\cite{anderson1991scheduler_activations}. These abstractions provide efficiency, isolation, and fairness for multi-process workloads. For emerging quantum computers, the underlying resources are qubits that are fragile, non-clonable, and coherence‐limited. \tool{} introduces new system abstractions and designs to re-establish practical resource sharing in this setting.

\paragraph{Qubit layout synthesis and routing.}
Quantum layout synthesis maps logical qubits and operations onto hardware with constrained connectivity, requiring efficient placement and routing to minimize SWAP overhead and mitigate noise. A rich body of work has optimized initial mapping and routing heuristics for single-circuit compilation~\cite{li2019tackling, murali2019noise, zhou2020quartz}. Tan and Cong~\cite{tan2020optimal} demonstrated significant optimality gaps in state-of-the-art mapping tools, with generated circuits frequently being $5$--$45\times$ deeper than optimal. Their findings underscore the importance of topology-aware compilation for maintaining fidelity in today’s hardware. Prior work assumes a single monolithic circuit and focuses on reducing qubit count through ancilla reuse within a circuit’s computation~\cite{sharma2025optimizing}. In contrast, \tool{} introduces a multi-process placement and routing strategy that allows helper qubits to be safely reassigned across concurrently executing circuits. This design introduces new constraints not captured in existing formulations, including inter-process crosstalk avoidance and prevention of unintended entanglement across workloads. To the best of our knowledge, \tool{} is the first system to enable inter-process helper-qubit reuse while preserving correctness and isolation guarantees.

\section{Conclusion}

{\color{red}}

We propose \tool{}, the first fine-grained resource sharing quantum operating system.  \tool{} significantly increases the throughput compared to the state-of-the-art approaches in helper qubit-intensive applications by 4.44$\times$, while maintains a competitive fidelity within 33\% of loss(Table \ref{tab:mainresult}).Our analysis of experiments data prove that the fidelity degradation primarily arises from the additional routing, reset operations, and helper-qubit reuse required at higher occupancy levels. Thus, \tool{} provides a fidelity throughput tradeoff and is adaptable for different users' requirements. 





\section*{Artifact}

We release our artifact and data in the anonymous repository: https://anonymous.4open.science/r/HALO-C814/.
\bibliographystyle{plain}
\bibliography{ref}

\end{document}